\documentclass{cambridge6A} 
\usepackage[numbers,square,comma]{natbib} 
\usepackage{rotating} 
\usepackage{floatpag}
\rotfloatpagestyle{empty} 
\usepackage{amsthm}
\usepackage{graphicx} 
\usepackage{multind}

\begin{document}                                              

\def\bE{\bar{E}}
\def\bR{\bar{R}}
\def\bu{\bar{u}}





\chapter{Observational Evidence for Black Holes}

\begin{center}  {Ramesh Narayan and Jeffrey E. McClintock \\
   ~~ \\
   Harvard Smithsonian Center for Astrophysics \\
   60 Garden Street, Cambridge, MA 02138, U.S.A.}
\end{center}

\begin{abstract}
Astronomers have discovered two populations of black holes:
(i)~stellar-mass black holes with masses in the range 5 to 30 solar
masses, millions of which are present in each galaxy in the universe,
and (ii)~supermassive black holes with masses in the range $10^6$ to
$10^{10}$ solar masses, one each in the nucleus of every galaxy.
There is strong circumstantial evidence that all these objects are
true black holes with event horizons.  The measured masses of
supermassive black holes are strongly correlated with properties of
their host galaxies, suggesting that these black holes, although
extremely small in size, have a strong influence on the formation and
evolution of entire galaxies. Spin parameters have recently been
measured for a number of black holes. Based on the data, there is an
indication that the kinetic power of at least one class of
relativistic jet ejected from accreting black holes may be correlated
with black hole spin. If verified, it would suggest that these jets
are powered by a generalized Penrose process mediated by magnetic
fields.
\end{abstract}

\section{Historical Introduction}
\label{intro}

\begin{figure}
    \includegraphics[scale=0.35]{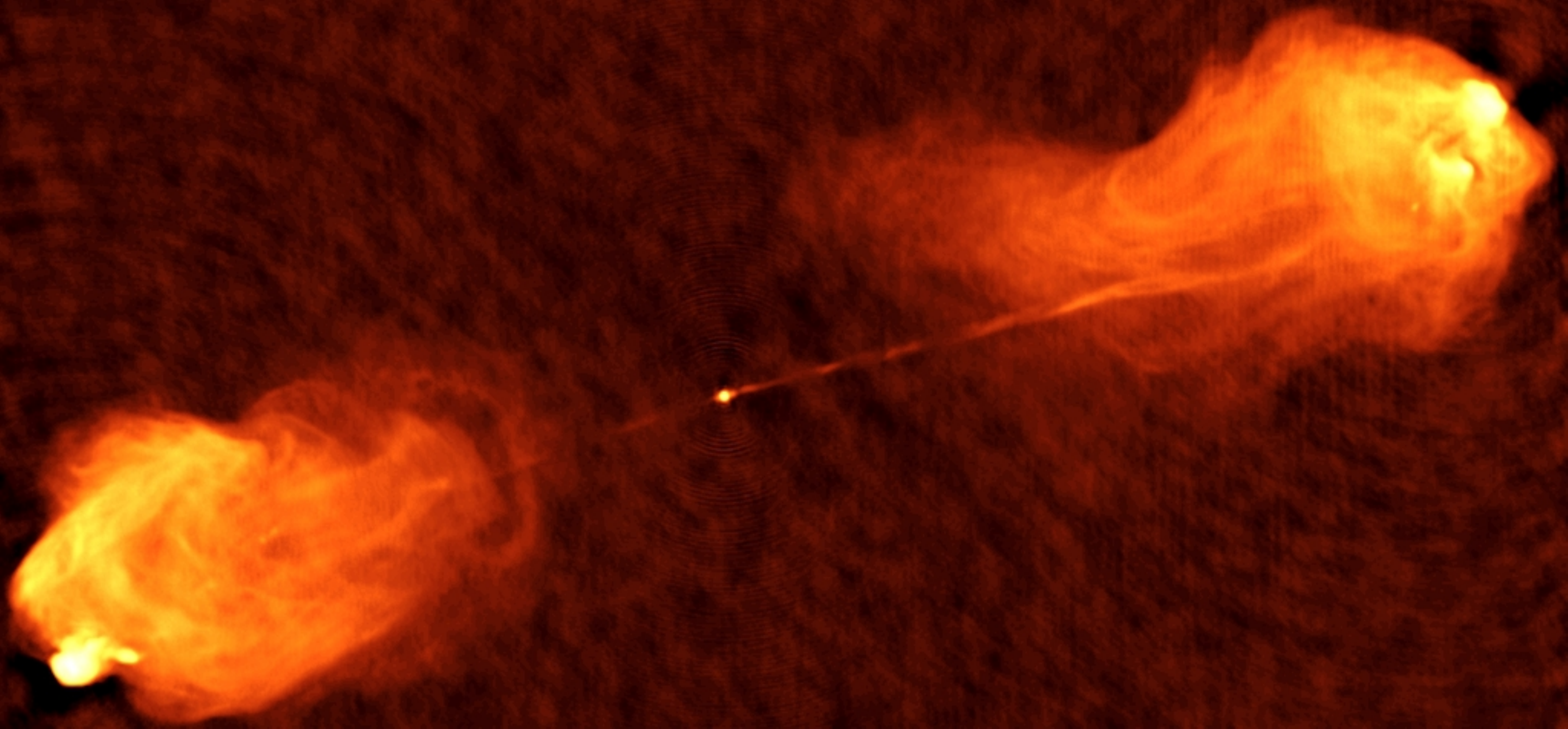}
    \caption
      {A modern radio image of Cyg A. The compact bright dot at the
        center of the image is the presumed supermassive black hole
        located in the nucleus of a giant elliptical galaxy. The two
        broken lines extending out on either side are relativistic
        jets carrying enormous amounts of energy in twin collimated
        beams. The jets are stopped by the intergalactic medium and
        then spread out into two giant lobes. All the observed radio
        emission is due to synchrotron radiation from relativistic
        electrons spiraling in magnetic fields. (Figure courtesy of
        C.~Carilli and R.~Perley, NRAO.)}
    \label{cyga}
  \end{figure}

The first astrophysical black hole to be discovered was Cygnus A,
which stood out already as a bright localized radio source in the
pioneering radio sky map of Grote Reber \citep{reb44,hey46}. The
recognition that Cyg A is a black hole, however, had to wait a few
decades. It required identifying the source with a distant galaxy
\citep{smi51,baa54}; resolving its radio image into a pair of radio
lobes \citep{jen53} with a compact source at the center \citep{har74}
(Fig.~\ref{cyga}); the discovery of quasars \citep{sch63}; and the
growing realization that all of these objects require an extremely
powerful but extraordinarily compact engine.  The only plausible
explanation is that Cyg A, like quasars and other active galactic
nuclei (AGN), is powered by a supermassive black hole.

The first stellar-mass black hole to be discovered was Cyg X--1 (also,
coincidentally, in the constellation of Cygnus), which was catalogued
in the early days of X-ray astronomy as a bright X-ray point source
\citep{bow65}. Evidence of its black hole nature came relatively soon.
The optical counterpart was confirmed to be a 5.6-day binary star
system in our Galaxy \citep{tan72}, and dynamical observations of the
stellar component showed that Cyg X--1 has a mass of at least several
solar masses \citep{bol72,web72}, making it too massive to be a
neutron star. It was therefore recognized as a black hole.

Cyg A and Cyg X--1 are members of two large but distinct populations
of black holes in the universe. We briefly review our current
knowledge of the two populations, and summarize the reasons for
identifying their members as black holes.

\section{Supermassive Black Holes}
\label{smbh}

\subsection{Mass and Size Estimates}
\label{smbhmass}

The radiation we see from supermassive black holes is produced by
accretion, the process by which gas spirals into the black hole from a
large radius. As the gas falls into the potential well, it converts a
part of the released potential energy to thermal energy, and
ultimately into radiation. A characteristic luminosity of any
gravitating object is the Eddington limit at which outward radiative
acceleration is balanced by the inward pull of gravity:
\begin{equation}
L_{\rm E} = \frac{4\pi GMm_pc}{\sigma_T} = 1.25\times
10^{38}\frac{M}{M_\odot} ~{\rm erg\,s^{-1}},
\end{equation}
where $M$ is the mass of the object, $M_\odot = 1.99\times 10^{33}$\,g
is the mass of the Sun, $m_p$ is the mass of the proton, and $\sigma_T
= 6.65\times10^{-25}\,{\rm cm^2}$ is the Thomson cross-section for
electron scattering.  A spherical object in equilibrium cannot have a
luminosity $L > L_{\rm E}$. Since bright quasars have typical
luminosities $L \sim10^{46}~{\rm erg\,s^{-1}}$, they must thus be very
massive: $M>10^8M_\odot$.

A large mass by itself does not indicate that an object is a black
hole. The second piece of information needed is its size.  In the case
of quasars and other AGN, a number of observations indicate that their
sizes are not very much larger than the gravitational radius of a
black hole of mass $M$:
\begin{equation}
R_g = \frac{GM}{c^2} = 1.48\times 10^5\frac{M}{M_\odot}\,{\rm cm}.
\end{equation}
The earliest indication for a small size came from the fact that
quasars show noticeable variability on a time scale of days. Since an
object cannot have large-amplitude variations on a time scale shorter
than its light-crossing time, it was deduced that quasars are no more
than a light-day across, i.e., their sizes must be $<10^2R_g$.

Modern limits are tighter. For instance, gravitational microlensing
observations of the quasar RXJ 1131--1231 indicate that the X-ray
emission comes from a region of size $\sim10R_g$ \citep{dai10}.
Tighter and more direct limits ($<$ few\,$R_g$) are obtained from
observations of the K$\alpha$ line of iron, which show that gas orbits
the central object at a good fraction of the speed of light
\citep{tan95,fab00}.  The only astrophysically plausible object
satisfying these mass and size limits is a supermassive black hole.

\subsection{The Mass of Sagittarius A$^*$}
\label{sgra}

Quasars are too distant for direct measurements of their mass.  The
situation is more favorable for supermassive black holes in the nuclei
of nearby galaxies. The majority of these black holes have very low
accretion luminosities and are thus very dim. This is an
advantage. Without the glare of a bright central source, it is
possible to carry out high resolution imaging and spectroscopic
observations relatively close to the black hole and thereby estimate
the black hole mass via dynamical methods. The most spectacular
results have been obtained in our own Milky Way Galaxy.

  \begin{figure}
    \includegraphics[scale=0.35]{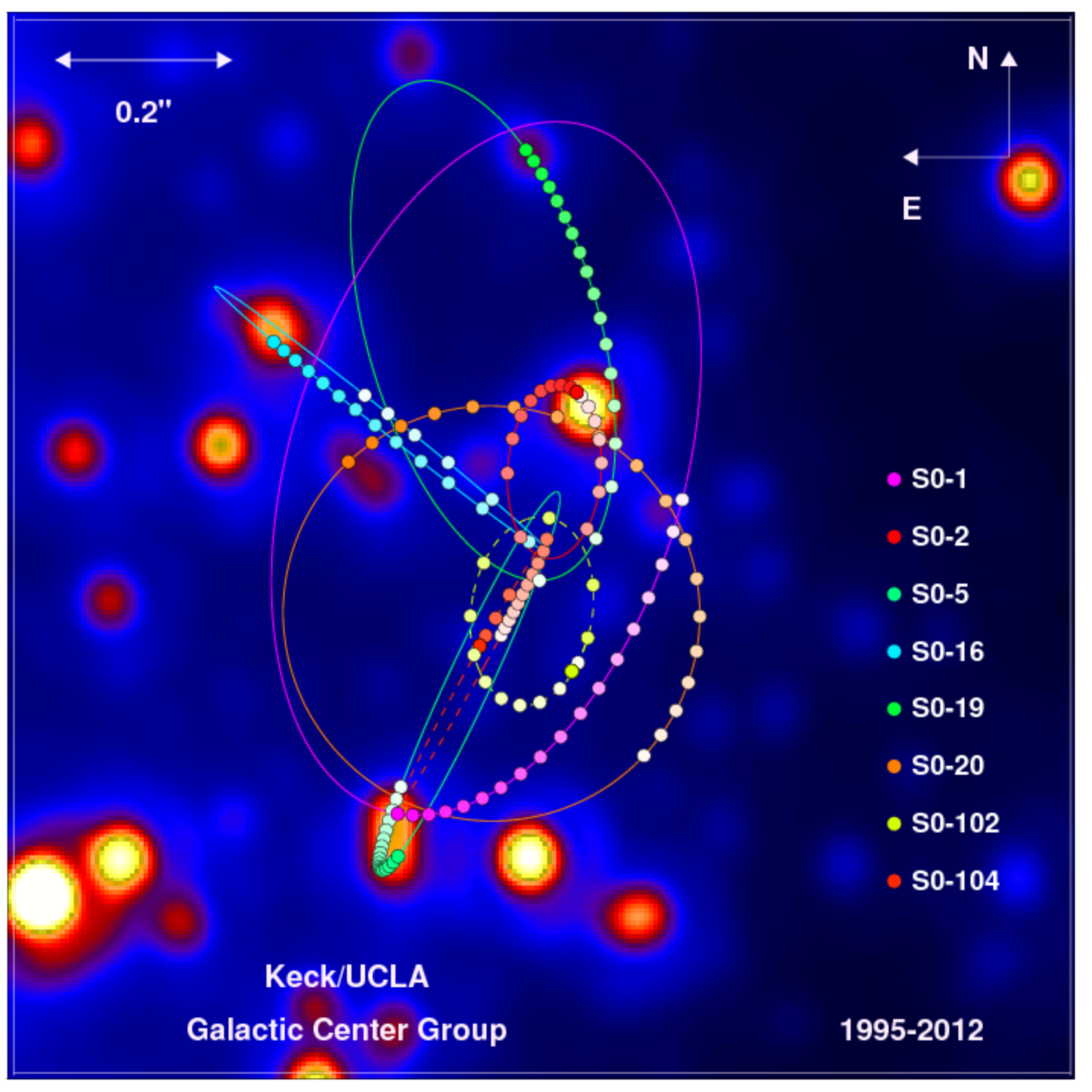}
    \caption
      {Orbital tracks on the plane of the sky over the period
        1995--2012 of 8 bright stars (S0--1, S0--2, ... , S0--104) at
        the center of our Galaxy. Keplerian fits to these orbits give
        the position and mass of the supermassive black hole in the
        nucleus of the Galaxy. (Figure courtesy of A.~Ghez and her
        research team at UCLA, based on data obtained with the
        W.~M.~Keck Telescopes.)}
    \label{Sgr}
  \end{figure}

Over the past twenty years, two different groups have successfully
used the largest telescopes on Earth to obtain diffraction-limited
infrared images of our Galactic Center, and have thereby mapped the
trajectories of stars orbiting the Galactic nucleus. Remarkably, all
the stars move on Keplerian orbits around a common focus
\citep{sch02,ghe05} (see Fig.~\ref{Sgr}) containing a dark mass of
$4.4\pm0.4\times10^6M_\odot$ \citep{mey12}.  Since the dark mass must
be interior to the pericenter of the most compact stellar orbit, its
radius must be $<10^3R_g$.

In fact, a much tighter limit can be placed on the size. At the center
of our Galaxy is a radio source called Sagittarius A$^*$ (Sgr
A$^*$). This source has been shown to be essentially at rest with
respect to the Galaxy, moving with a speed less than about $1\,{\rm
  km\,s^{-1}}$ \citep{rei04}. Given the huge velocities of stars
orbiting in its vicinity (the fastest stars mentioned in the previous
paragraph move with speeds up to $10^4~{\rm km\,s^{-1}}$),
equipartition arguments imply that Sgr A$^*$ must be more massive than
$10^5M_\odot$.  The only plausible interpretation is that Sgr A$^*$ is
identical to the $4.4\times10^6M_\odot$ object inferred from stellar
orbits.

Meanwhile, Sgr A$^*$ is a bright radio source and has been imaged with
exquisite precision using interferometric techniques. The most recent
observations indicate that emission at 1.3\,mm wavelength comes from
within a radius of a few $R_g$ \citep{doe08,fis11}. This robust size
constraint makes it virtually certain that Sgr A$^*$ must be a
supermassive black hole.

\subsection{Other Nearby Supermassive Black Holes}
\label{othersmbh}


Occasionally, the orbiting gas in the accretion disk around a
supermassive black hole produces radio maser emission from transitions
of the water molecule. If the galaxy is sufficiently nearby, the maser
emitting ``blobs'' can be spatially resolved by interferometric
methods and their line-of-sight velocities can be measured accurately
by the Doppler technique. The most spectacular example is the nearly
edge-on disk in the nucleus of the galaxy NGC~4258
\citep{miy95,gre95}, where the measured velocities follow a perfect
Keplerian profile.
The required black hole mass is $4.00\pm0.09\times10^7M_\odot$
\citep{hum13}.

Maser disks are relatively rare. A more widely used method employs
high spatial resolution observations in the optical band with the
Hubble Space Telescope (see \citep{kor13} for a comprehensive review).
By simultaneously fitting the spatial brightness distribution and
two-dimensional line-of-sight velocity distribution of stars in the
vicinity of a galactic nucleus, and using advanced three-integral
dynamical models for stellar orbits, it is possible to estimate the
mass of a compact central object, if one is present. Several tens of
black hole masses have been measured by this method, with
uncertainties typically at about a factor of two. In the majority of
cases, models without a compact central mass are ruled out with high
statistical significance.  From these studies it has become clear that
essentially every galaxy in the universe hosts a supermassive black
hole in its nucleus.

Other less direct methods are available for measuring masses of more
distant black holes. Two methods in particular, one based on
reverberation mapping \citep{pet04} and the other on an empirical
linewidth-luminosity relation \citep{ves06}, deserve mention.

\subsection{A Remarkable Correlation}
\label{Msigmacorr}

  \begin{figure}
\vskip -2in
    \includegraphics[scale=0.6]{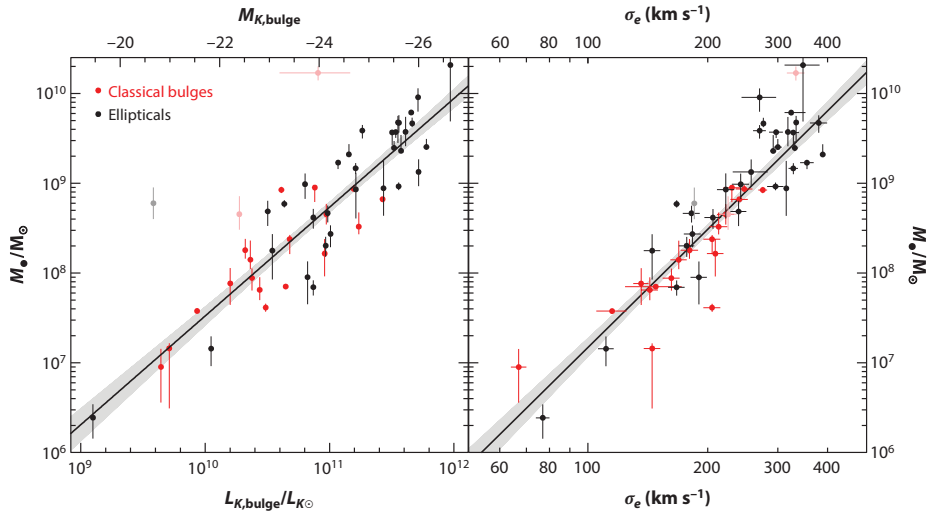}
\vskip -2in
    \caption
      {Observed correlations between supermassive black hole mass
        $M_\bullet$ and (left) the infrared luminosity of the bulge of
        the host galaxy in units of solar luminosities (represented in
        the top axis by the absolute magnitude $M_{\rm K,bulge}$), and
        (right) the velocity dispersion $\sigma_e$ of the stars in the
        bulge. (Reprinted with permission from \citep{kor13}.)}
    \label{Msigma}
  \end{figure}

Unquestionably, the most dramatic discovery to come out of the work
described in the previous subsection is the fact that supermassive
black hole masses are correlated strongly with the properties of their
host galaxies. Figure~\ref{Msigma} (from \citep{kor13}) shows two
such correlations: (a) between the black hole mass $M$ and the
luminosity (in this case infrared luminosity) of the bulge of the
galaxy \citep{mag98}, and (b) between $M$ and the stellar velocity
dispersion $\sigma_e$ of the bulge \citep{fer00,geb00}.

At first sight, these correlations are baffling. The mass of the black
hole is typically $10^3$ times smaller than that of the bulge, and its
size ($R_g$) is $10^8$ times smaller. How could such an insignificant
object show such a strong correlation with its parent galaxy?  The
answer can be understood by considering a more relevant quantity than
either mass or radius alone: the binding energy $GM^2/R$.  In terms of
binding energy, the black hole is actually ``stronger'' (by quite a
bit) than the entire galaxy. Indeed, the current paradigm, and a major
area of research, is that supermassive black holes exert considerable
``feedback'' on their host galaxies during the formation and growth of
both entities. As a natural consequence, their parameters become
strongly correlated in the manner shown in Fig.~\ref{Msigma}
(e.g.~\citep{dim05,kin03,kin05}).

As an important practical application, the above correlations can be
used to estimate the masses of high redshift black holes using the
luminosities or stellar velocity distributions of their host galaxies.

\section{Stellar-Mass Black Holes}\label{sec:stellarBHs}
\label{stellarbh}

\begin{figure}
    \includegraphics[scale=0.55]{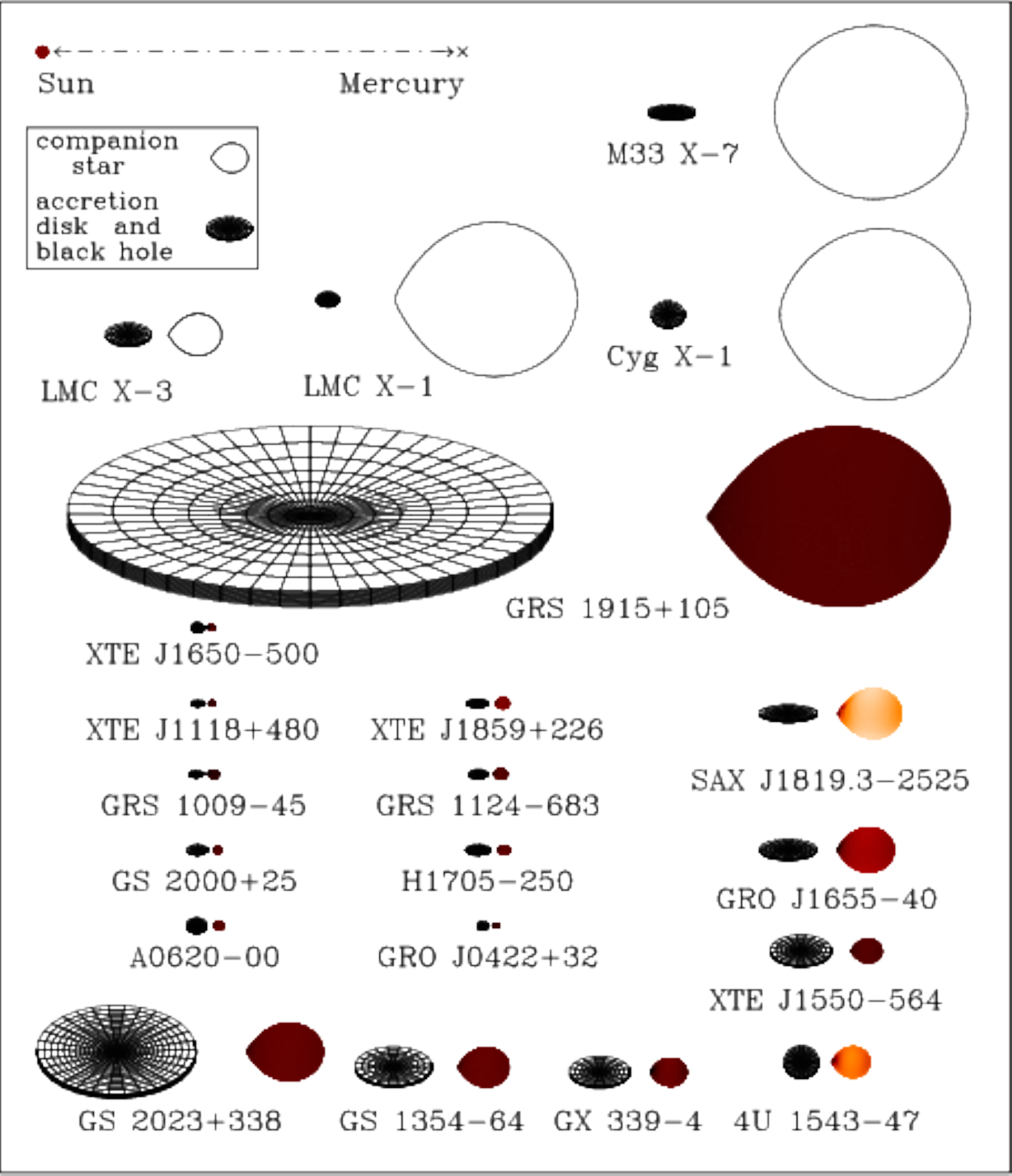}
    \caption
      {Sketches of 21 black hole binaries (see scale and legend in the
        upper-left corner).  The tidally-distorted shapes of the
        companion stars are accurately rendered in Roche geometry.
        The black holes are located at the centers of the disks.  A
        disk's tilt indicates the inclination angle $i$ of the binary,
        where $i=0$ corresponds to a system that is viewed face-on;
        e.g., $i=21^{\circ}$ for 4U 1543--47 (bottom right) and $i =
        75^{\circ}$ for M33 X--7 (top right).  The size of a system is
        largely set by the orbital period, which ranges from 33.9 days
        for the giant system GRS~1915+105 to 0.2 days for tiny XTE
        J1118+480. Three systems hosting persistent X-ray sources ---
        M33 X--7, LMC X--1 and Cyg X--1 --- are located at the top.  The
        other 18 systems are transient sources.  (Figure courtesy of
        J.~Orosz.)}
    \label{fig:21bhbs}
  \end{figure}

Many millions of stellar-mass black holes are inferred to be present
in our Milky Way galaxy, and in every other galaxy in the universe,
but the existence of only 24 of them has been confirmed via dynamical
observations.  These 24, whose masses are in the range $M =
5-30~M_{\odot}$, are located in X-ray binary systems, 21 of which are
sketched to scale in Fig.~\ref{fig:21bhbs}.  The X-rays are produced
by gas that flows from the companion star on to the black hole via an
accretion disk. Close to the black hole (radii $\sim10R_{g}$), the
accreting gas reaches a typical temperature of $\sim10^7$~K and emits
X-rays.

The 24 black hole binaries divide naturally into two classes: (i) 5 of
these black holes are {\it persistent} X-ray sources, which are fed by
winds from their massive companion stars. (ii) The remaining 19
binaries are {\it transient} sources, whose X-ray luminosities vary
widely, ranging from roughly the Eddington luminosity $L_E$ down to as
low as $\sim10^{-8}L_E$.  A typical transient black hole is active for
about a year and then remains quiescent for decades.

\subsection{Mass Measurements}\label{sec:mass}
\label{stellarmass}

The masses of stellar-mass black holes are measured by employing the
same methodologies that have been used for over a century to measure
the masses of ordinary stars in binary system.  Most important is the
radial velocity curve of the companion star, which is derived from a
collection of spectroscopic observations that span an orbital cycle.
These velocity data deliver two key parameters: the orbital period $P$
and the semi-amplitude of the velocity curve $K$, which in turn
determine the value of the mass function: 
\begin{equation}
f(M) \equiv \frac{{P}K^{3}}{2\pi G} = \frac{M\sin^3i}{(1+q)^{2}},
\end{equation}  
where $i$ is the orbital inclination angle of the binary
(Fig.~\ref{fig:21bhbs}) and $q$ is the ratio of the companion star
mass to that of the black hole.

An inspection of the above equation shows that the value of the
observable ${P}K^{3}/2\pi G$, which can be accurately measured, is the
absolute minimum mass of the black hole.  For ten out of the total
sample of 24 stellar-mass black holes, this minimum mass ranges from
$3-8~M_{\odot}$ (see Table 2 in \citep{oze10} and \citep{stg13}). In
comparison, the maximum stable mass of a neutron star is widely agreed
to be less than about $3~M_{\odot}$ \citep{rho74,kal96}.  Therefore,
on the basis of a single robust observable, one can conclude that
these ten compact X-ray sources must be black holes.

In order to obtain the actual masses of these and other stellar-mass
black holes, one must additionally determine $q$ and $i$.  The mass
ratio $q$ is usually estimated by measuring the rotational velocity of
the companion star.  The inclination angle $i$ can be constrained in
several ways; commonly, one models the light curve of the
tidally-distorted companion star.  Selected mass measurements for
seven black holes are given in Table~\ref{tab:data}.  For further
details on measuring the masses of black holes, see the references
cited in the table and \citep{cha06}.

\begin{table*}
\label{mass-spin}
\caption{Masses and spins, determined via the continuum-fitting
  method, for a selected sample of seven black holes.  By the No-Hair
  Theorem (\S\ref{nohair}), the data constitute complete descriptions of
  these black holes$^a$.
\label{tab:data}}
\footnotesize
\begin{tabular}{lccl}
\hline
\noalign
{\vspace{1mm}}
System & $M/M_{\odot}$ & $a_*$ & References \\
\noalign
{\vspace{-2.5mm}}
&&& \\
\hline
\noalign
{\vspace{1.2mm}}
Persistent &&& \\
\noalign
{\vspace{1mm}}
\hline
\noalign
{\vspace{1.2mm}}
Cyg X--1 & $14.8\pm1.0$ & $> 0.95$ & \citep{oro11a}; \citep{gou11} \\
{\vspace{-3.5mm}}
&&& \\
\noalign
{\vspace{0.9mm}}
LMC X--1 & $10.9\pm1.4$ & $0.92_{-0.07}^{+0.05}$ & \citep{oro09}; \citep{gou09} \\
\noalign
{\vspace{0.9mm}}
M33 X--7 & $15.65\pm1.45$ & $0.84\pm0.05$ & \citep{oro07}; \citep{liu08} \\
\noalign
{\vspace{1mm}}
\hline
\noalign
{\vspace{1mm}}
Transient &&& \\
\noalign
{\vspace{1mm}}
\hline
\noalign
{\vspace{1.2mm}}
GRS 1915+105 & $10.1\pm0.6$ & $> 0.95^b$ & \citep{stg13}; \citep{mcc06} \\
{\vspace{-3.5mm}}
&&& \\
\noalign
{\vspace{0.9mm}}
GRO J1655--40 & $6.3\pm0.5$ & $0.70\pm0.10^b$ & \citep{gre01}; \citep{sha06} \\
\noalign
{\vspace{0.9mm}}
XTE J1550--564 & $9.1\pm0.6$ & $0.34_{-0.28}^{+0.20}$ & \citep{oro11b}; \citep{ste11} \\
\noalign
{\vspace{0.9mm}}
A0620--00 & $6.6\pm0.25$ & $0.12\pm0.19$ & \citep{can10}; \citep{gou10} \\ 
\noalign
{\vspace{0.7mm}}
\hline
\end{tabular}
\vskip 2mm
Notes: \newline
$^a$~Errors are quoted at the 68\% level of confidence, except for the
two spin limits, which are estimated to be at the 99.7\% level of confidence. \newline
$^b$~Uncertainties are greater than those in papers cited because early error estimates were crude. \newline
\end{table*}

\subsection{Spin Estimates}\label{sec:spin}
\label{spin}

Spin is difficult to measure because its effects manifest only near
the black hole ($R<10R_g$).  One must not only make discerning
observations in this tiny region of space-time, but one must also have
a reliable model of the accretion flow in strong gravity.  Two
fortunate circumstances come to the rescue: (i) We do have at least
one simple black hole accretion model, viz., the thin accretion disk
model \citep{sha73,nov73}.  (ii) Among the several distinct and
long-lived accretion states observed in individual stellar-mass black
holes \citep{rem06}, one particularly simple state, called the {\it thermal}
state, is dominated by emission from an optically thick accretion
disk, and is well-described by the thin disk model.

According to the thin disk model, the inner edge of the accretion disk
is located at the radius of the innermost stable circular orbit
$R_{\rm ISCO}$. Moreover, $R_{\rm ISCO}/R_g$ is a monotonic function
of the dimensionless black hole spin parameter $a_* = cJ/GM^2$
\citep{bar72}, where $J$ is the angular momentum of the black hole
(note, $|a_*|<1$).  In the continuum-fitting method of measuring spin
\citep{zha97,sha06,mcc06}, one observes radiation from the accreting
black hole when it is in the thermal state. One then estimates $R_{\rm
  ISCO}$, and hence $a_*$, by fitting the thermal continuum spectrum
to the thin-disk model. The method is simple: It is strictly analogous
to using the theory of blackbody radiation to measure the radius of a
star whose flux, temperature and distance are known.  By this analogy,
it is clear that to measure $R_{\rm ISCO}$ one must measure the flux
and temperature of the radiation from the accretion disk, which one
obtains from X-ray observations. One must also measure the source
distance $D$ and the disk inclination $i$ (an extra parameter that is
not needed for a spherical star).  Additionally, one must know $M$ in
order to scale $R_{\rm ISCO}$ by $R_g$ to determine $a_*$. The
uncertainties in all these ancilliary measurements contribute to the
overall error budget.

The spins of ten black holes have been measured by the
continuum-fitting method, seven of which are presented in
Table~\ref{tab:data}.  The robustness of the method is demonstrated by
the dozens or hundreds of independent and consistent measurements of
spin that have been obtained for several black holes, and through
careful consideration of many sources of systematic error.  For a
review of the continuum-fitting method and a summary of results, see
\citep{mcc13}.

An alternative method of measuring black hole spin, in which one
determines $R_{\rm ISCO}/R_g$ by modeling the profile of the broad and
skewed fluorescence Fe K$\alpha$ line, has been widely practiced since
its inception \citep{fab89}.  However, obtaining reliable results for
stellar-mass black holes is challenging because one must use data in
states other than the thermal state, where the disk emission is
strongly Comptonized and harder to model. Furthermore, the basic
geometry of the disk is poorly constrained, and it is even doubtful
that the inner edge of the disk is located at $R_{\rm ISCO}$.

To date, the Fe-line method has been used to estimate the spins of
more than a dozen stellar-mass black holes. A few of these black holes
have been studied using both the continuum-fitting and Fe-line
methods, and there is reasonable agreement between the two independent
spin estimates.  The Fe-line method is especially important in the
case of supermassive black holes \citep{rey13}, where it is difficult
to apply the continuum-fitting method.

\subsection{Intermediate Mass Black Holes}\label{sec:imbhs}
\label{imbh}

Are there black holes of intermediate mass, i.e., black holes that are
too massive ($M > 100~M_{\odot}$) to have formed from ordinary stars
but, at the same time, are not in the nucleus of a galaxy?  Such
objects, referred to as intermediate mass black holes (IMBHs), would
represent a new and distinct class of black hole.  The leading IMBH
candidates are the brightest ``ultraluminous'' X-ray sources in
external galaxies, whose observed luminosities can be up to
$\sim100-1000$ times the Eddington luminosity of a $10 M_{\odot}$
black hole.  Although there are some promising candidates
(e.g.~\citep{fen10,dav11}), none has been confirmed because of the
difficulties of obtaining a firm dynamical measurement of mass.

\section{Physics of Astrophysical Black Holes}
\label{physics}

\subsection{Are They Really Black Holes?}
\label{eventhorizon}

The astrophysical black holes discussed so far are technically only
black hole candidates. True, they are sufficiently massive and compact
that we cannot match the observations with any object in stable
equilibrium other than a black hole. However, this by itself does not
prove that the objects are true black holes, defined as objects with
event horizons.  Black hole candidates could, in principle, be exotic
objects made of some kind of unusual matter that enables them to have
a surface (no horizon), despite their extreme compactness.

Astronomers have devised a number of tests to check whether black hole
candidates have a ``surface''. In brief, all the evidence to date
shows that black hole candidates do not have normal surfaces that are
visible to distant observers (see \citep{nar08,bro09}; and
references therein). The arguments are sufficiently strong that ---
barring scenarios that are more bizarre than a black hole --- they
essentially ``prove'' that the astrophysical black hole candidates
discussed in this article possess event horizons. However, the proof
is still indirect \citep{abr02}.

\subsection{Spinning Black Holes and the Penrose Process}
\label{penrose}

It is a remarkable consequence of black hole theory that a spinning
black hole has free energy available to be tapped.  \citet{pen69}
showed via a simple toy model that particles falling into a spinning
hole on negative energy orbits can extract some of the black hole's
spin energy. Energy extraction is allowed by the Area Theorem, which
states that the horizon area $A$ of a black hole cannot decrease with
time: $dA/dt \geq 0$ (e.g., \citep{bar73}). For a black hole of mass
$M$ and dimensionless spin $a_*$, the area is given by $A=8\pi R_g^2\,
[1+(1-a_*^2)^{1/2}]$. Therefore, a black hole can lose energy and
mass, and thus reduce the magnitude of $R_g$, provided the spin
parameter $a_*$ also decreases by a sufficient amount to satisfy the
Area Theorem.  In effect, the hole spins down and gives up some of its
spin-energy to infinity.  Penrose's negative energy particles are a
conceptually transparent way of demonstrating this effect.

An astrophysically more promising scenario for the extraction of spin
energy makes use of magnetized accretion flows, as outlined in early
papers \citep{ruf75,bla77}.  In this mechanism, the
dragging of space-time by a spinning hole causes magnetic field lines
to be twisted, resulting in an outflow of energy and angular momentum
along field lines. Does this actually happen anywhere in the universe?
Astronomers have long hypothesized that relativistic jets such as
those in Cygnus~A (Fig.~\ref{cyga}) might be explained by some such
process.

In recent years, general relativistic magnetohydrodynamic (MHD)
numerical simulations of accreting spinning black holes have been
carried out that show relativistic jets forming naturally from fairly
generic initial conditions.  More importantly, the simulations show
unambiguously that, in at least some cases, the jet receives its power
from the spin energy of the black hole and not from the accretion disk
\citep{tch11,pen13,las13}. Specifically, energy and angular momentum
flow directly from the black hole through the jet to the external
universe, and the mass and spin of the black hole consequently
decrease.  The jet power varies approximately as $a_*^2$, and is thus
largest for the most rapidly spinning holes. In brief, a generalized
MHD version of the Penrose process operates naturally and efficiently
in idealized simulations on a computer.

  \begin{figure}
    \includegraphics[scale=0.35]{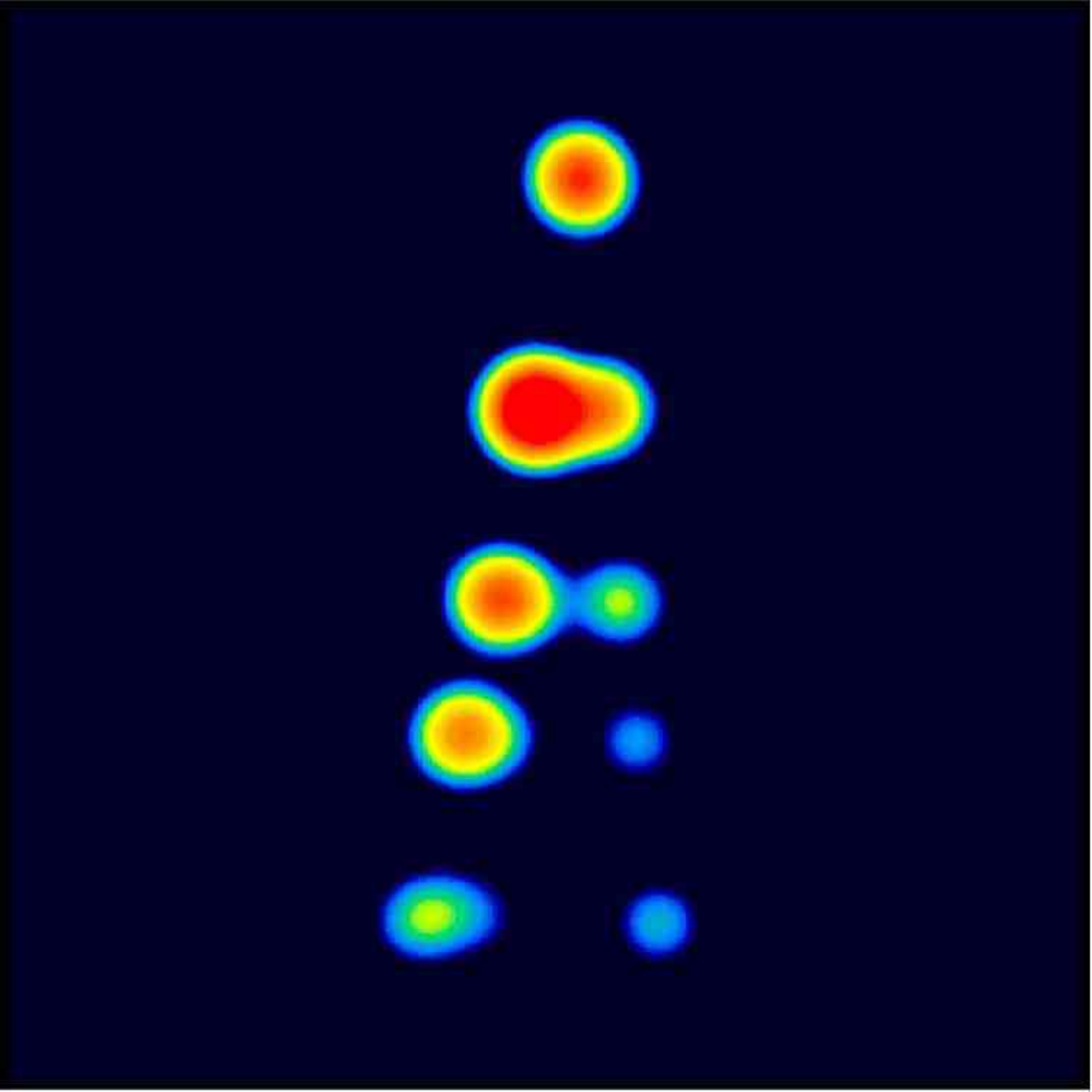}
    \caption
      {A sequence of radio images \citep{mir94} of the transient black
        hole X-ray binary GRS~1915+105 during the period 18 March 1994
        (uppermost image consisting of a single blob) to 16 April 1994
        (lowermost image with two widely separated blobs). Two
        radio-emitting blobs were ejected from the source around the
        time of the first observation, and they subsequently moved
        ballistically outward from the source. The blob on the left
        has an apparent speed on the sky greater than the speed of
        light (superluminal motion), which is a relativistic
        effect. The Lorentz factor of each blob is estimated to be
        $\gamma\approx2.6$. (Figure courtesy of F.~Mirabel.)}
    \label{1915}
  \end{figure}

  \begin{figure}
    \includegraphics[scale=0.5]{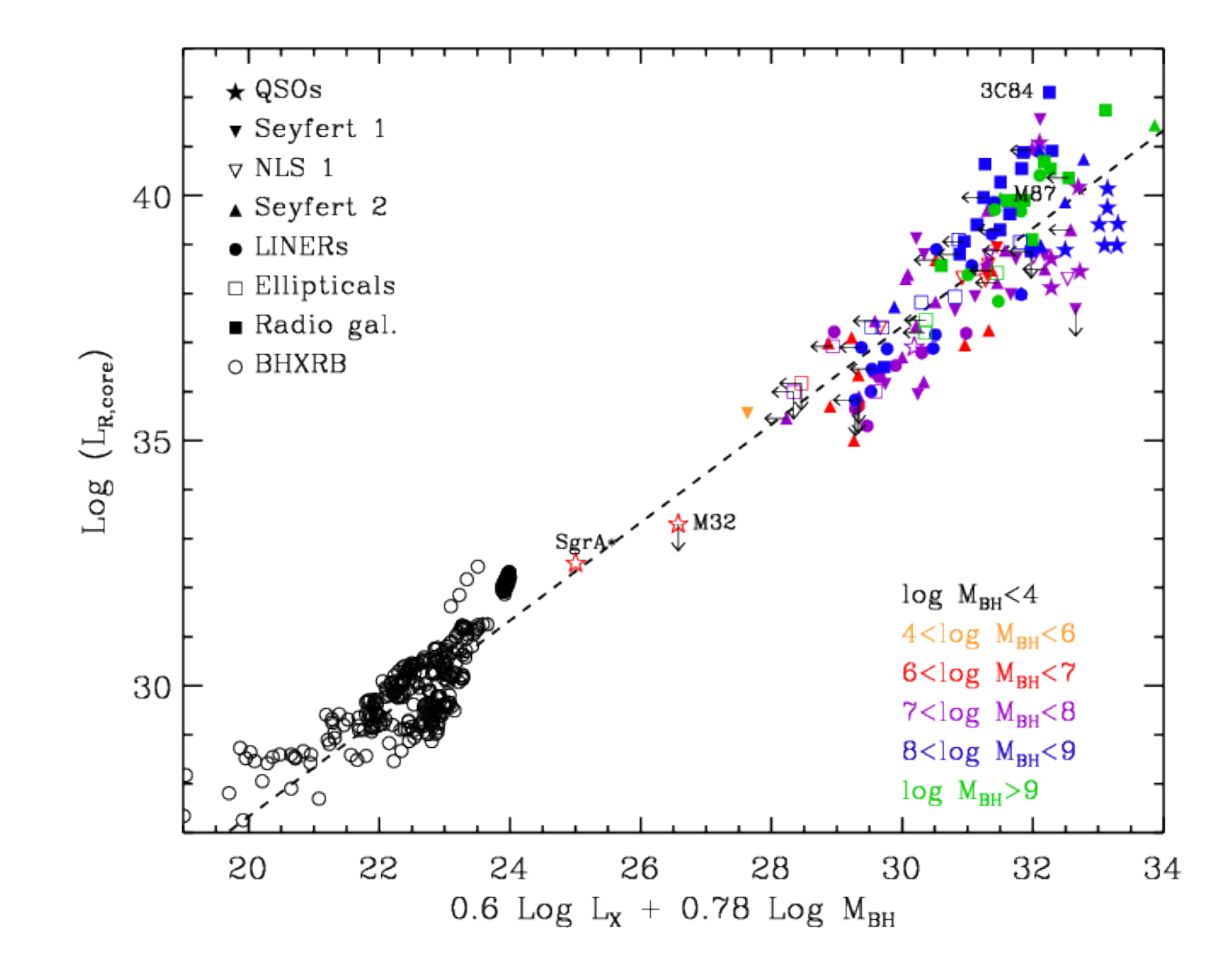}
    \caption
       {The ``fundamental plane of black hole activity''
         \citep{mer03}.  The correlation extends over many decades of
         black hole mass and accretion luminosity, and includes many
         different source types. (Figure courtesy of A.~Merloni.)}
    \label{fundplane}
  \end{figure}

The observational situation is less clear. Radio-emitting relativistic
jets have been known in AGN for many decades (e.g., Fig.~\ref{cyga}),
and more recently, jets have been discovered also in stellar-mass
black holes (Fig.~\ref{1915} shows a famous example). A very
interesting relation has been found between the radio luminosity
$L_R$, which measures jet power, the X-ray luminosity $L_X$, which
measures accretion power, and the black hole mass $M$
\citep{mer03,hei03,fal04}. This relation, called the ``fundamental
plane of black hole activity'' (Fig.~\ref{fundplane}), extends over
many decades of the parameters, connecting the most massive and
luminous AGN with stellar-mass black holes. The relatively tight
correlation, which is further emphasized in recent work \citep{van13},
implies that jet power depends primarily on the black hole mass and
accretion rate, leaving little room for an additional dependence on
black hole spin. However, most of the black holes plotted on the
fundamental plane do not have spin measurements, so the argument for a
lack of spin-dependence is somewhat indirect.

\begin{figure}
  \includegraphics[scale=0.26]{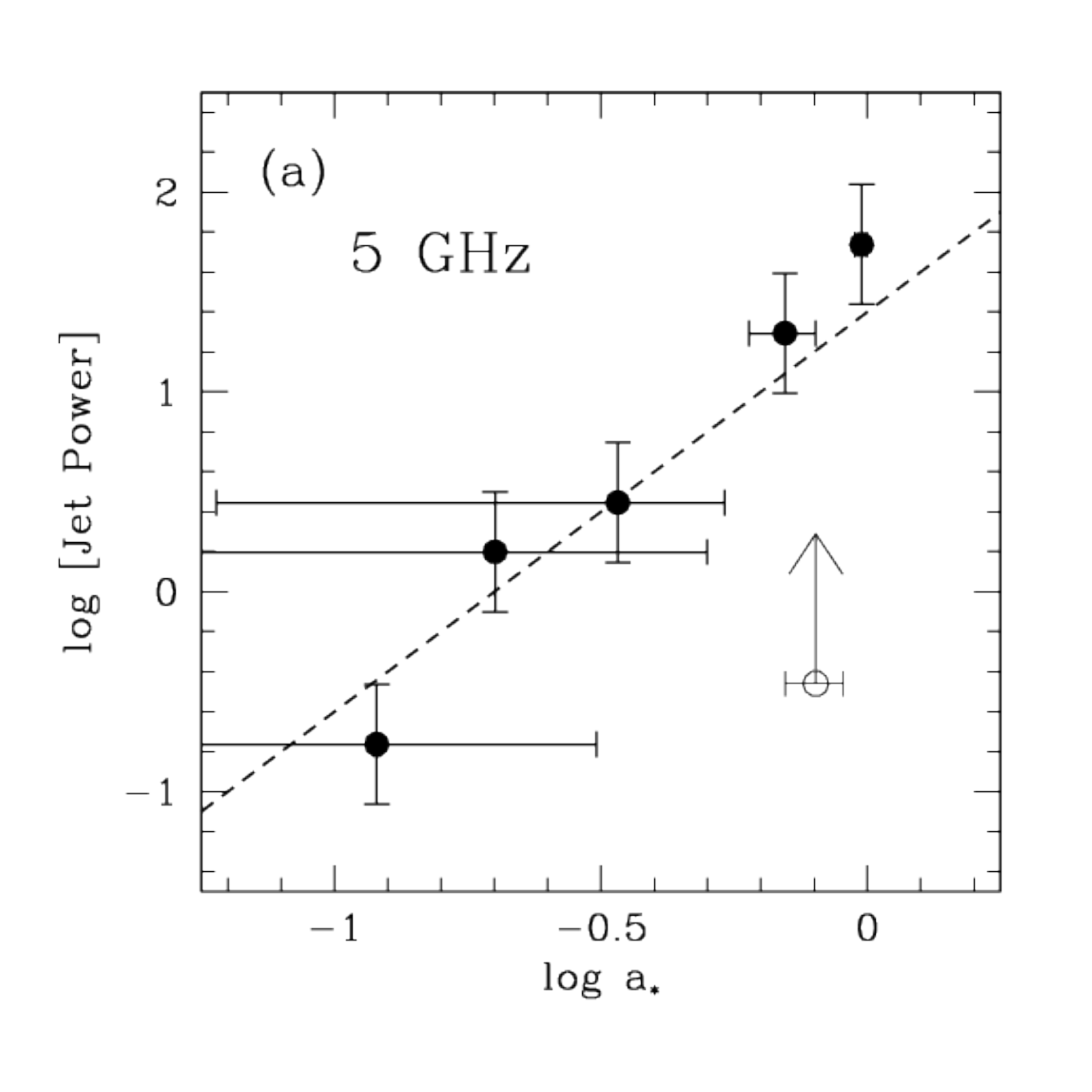}
  \includegraphics[scale=0.30,angle=90]{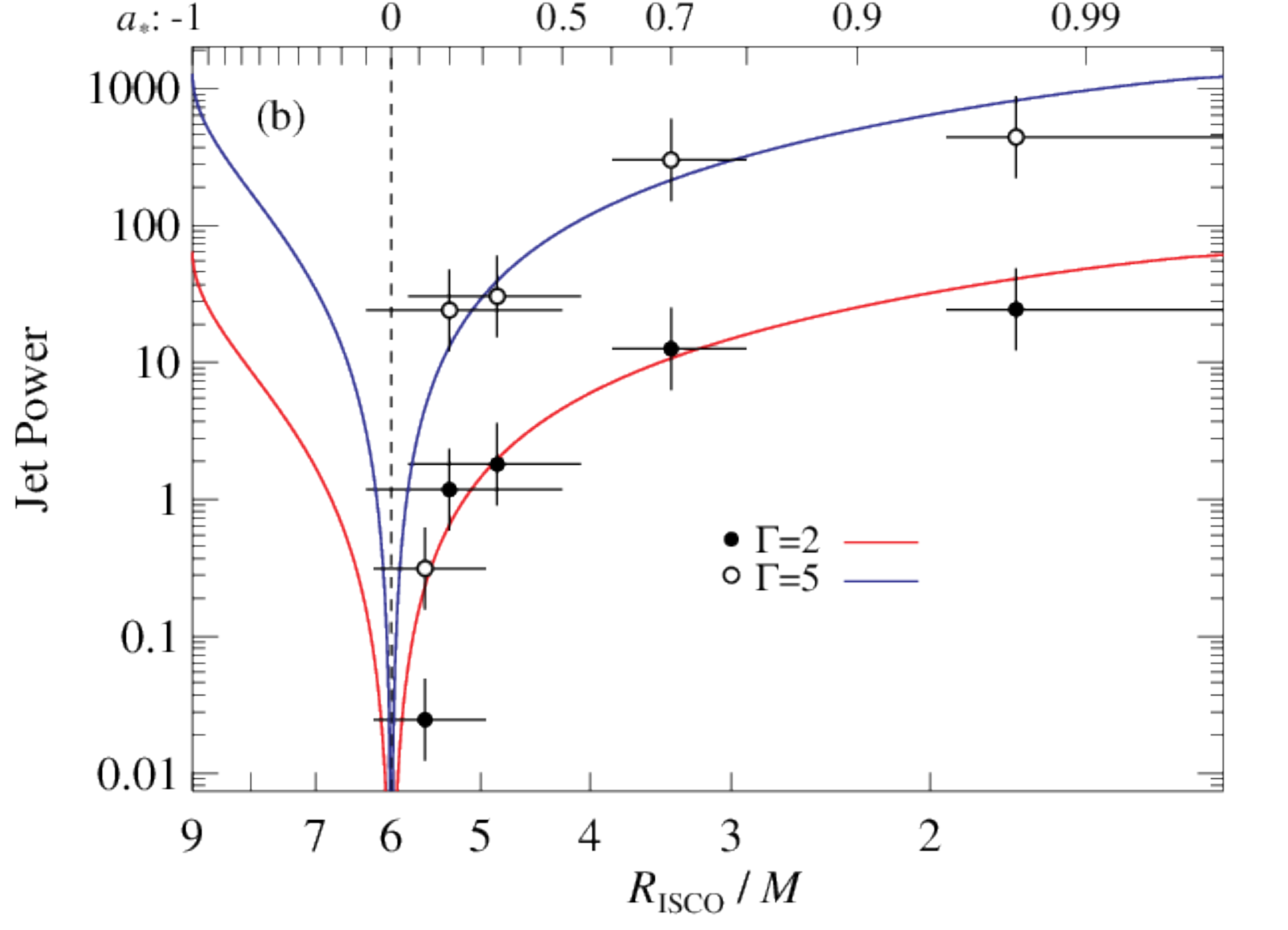}
\caption{(a) Plot of jet power, estimated from 5\,GHz radio flux at
  light curve maximum, versus black hole spin, measured via the
  continuum-fitting method, for five transient stellar-mass black
  holes \citep{nar12,ste13}. The dashed line has slope fixed to 2 (as
  predicted by theoretical models \citep{bla77}) and is not a
  fit. (b) Plot of jet power versus $R_{\rm ISCO}/R_g$.  Here jet
  power has been corrected for beaming assuming jet Lorentz factor
  $\Gamma=2$ (filled circles) or $\Gamma=5$ (open circles).  The two
  solid lines correspond to fits of a relation of the form ``Jet
  Power'' $\propto \Omega_H^2$, where $\Omega_H$ is the angular
  frequency of the black hole horizon. Note that the jet power varies
  by a factor of $\sim10^3$ among the five objects shown.}
\label{fig:jet}
\end{figure}

Meanwhile, tentative but direct observational evidence for a
correlation between black hole spin and jet power has been found in a
sample of stellar-mass black holes for which spins had been previously
measured \citep{nar12,ste13} (Fig.~\ref{fig:jet}). The evidence is
still controversial \citep{rus13,mcc13}, in large part because of the
small size of the sample. In addition, the correlation is restricted
to stellar-mass black holes that accrete at close to the Eddington
limit and produce so-called ``episodic'' or ``ballistic'' jets (e.g.,
Fig.~\ref{1915}), which are different from the jets considered for the
fundamental plane (for a discussion see \citep{fen04}).

In summary, theory and numerical simulations suggest strongly that
relativistic jets are powered by black hole spin, i.e., by a
generalized Penrose process. However, observational evidence is
limited to that shown in Fig.~\ref{fig:jet} and is in comparison weak.

\subsection{Testing the No-Hair Theorem}
\label{nohair}

The No-Hair Theorem states that stationary black holes, such as those
discussed herein, are completely described by the Kerr metric, which
has only two parameters: black hole mass $M$ and spin parameter
$a_*$.\footnote{In principle, a black hole can have a third parameter,
  electric charge, but the black holes studied in astrophysics are
  unlikely to have sufficient charge to be dynamically important.}
Testing this theorem requires measuring $M$ and $a_*$ of a black hole
with great accuracy and demonstrating that no additional parameter is
needed to explain any observable. At the present time, mass and spin
measurements of stellar-mass and supermassive black holes are not
accurate enough, nor are there a sufficient number of independent
observables, to permit such a test.

The most promising system for testing the No-Hair Theorem is Sgr
A$^*$, the supermassive black hole in our Galaxy
(\S\ref{sgra}). Within the next decade, ultra-high resolution
interferometric observations are planned at millimeter wavelengths
with the ``Event Horizon Telescope'' \citep{doe09}, which will produce
direct images of the accreting gas in Sgr A$^*$ on length scales
comparable to the horizon.  These measurements could potentially be
used to test the No-Hair Theorem (e.g.~\citep{joh10}).

\section{Conclusion}
\label{summary}

The dawning that black holes are real occurred at the midpoint of this
century of General Relativity, at the First Texas Symposium on
Relativistic Astrophysics in 1963.  There, Roy Kerr announced his
solution, Jesse Greenstein described Maarten Schmidt's discovery of
quasars, and Harlan Smith reported on the rapid variability of these
objects \citep{sch89}.  Today, black hole astrophysics is advancing at
a breathtaking rate.  Tomorrow, spurred on by the commissioning of the
Event Horizon Telescope and the advent of gravitational wave
astronomy, it is reasonable to expect the discovery of many new
unimaginable wonders.

\end{document}